\documentstyle[aps,twocolumn,epsf]{revtex}
\newtheorem{twist}{Twist Theorem}
\begin{document}


\title{Bi$_2$Sr$_2$CaCu$_2$O$_{8+\delta}$ Bicrystal $c$-Axis Twist Josephson
Junctions:  A New Phase-Sensitive Test of Order Parameter Symmetry}

\author{Richard A. Klemm}

\address{Max-Planck-Institut f{\"u}r Physik komplexer Systeme,
N{\"o}thnitzer Stra{\ss}e 38, D-01187 Dresden, Germany}


\maketitle

\begin{abstract}
Li {\it et al.} [Phys. Rev. Lett. {\bf 83}, 4160 (1999)] prepared atomically clean Bi$_2$Sr$_2$CaCu$_2$O$_{8+\delta}$
(BSCCO) Josephson junctions between identical single crystal cleaves stacked and
twisted an angle $\phi_0$ about the $c$-axis.  For each bicrystal, the ratio
$J_c^J/J_c^S$ of the $c$ axis twist junction critical current density to that
across either single crystal part is unity, independent of $\phi_0$ and the
ratio $A^J/A^S$ of junction areas.  From extensive theoretical studies
involving a variety of tunneling and superconducting order parameter (OP) forms, 
we conclude that
the results provide strong evidence for incoherent $c$-axis tunneling and that
the dominant OP is $s$-wave for $T\le
T_c$. Recently,  Takano {\it et al.} [Phys. Rev. B {\bf
65}, 140513(R) (2002)]  obtained results from BSCCO whisker
twist junctions which also rule out a pure $d$-wave OP, but which are
surprisingly suggestive of coherent $c$-axis tunneling from small Fermi
surface hot spots.  
\end{abstract}

\section{Introduction}

Phase-sensitive
experiments to test the symmetry of the
superconducting order parameter (OP) in the high transition temperature ($T_c$)
superconductors  were mostly made on 
YBa$_2$Cu$_3$O$_{7-\delta}$ (YBCO) \cite{tk}, for which the OP can have  mixed ($d_{x^2-y^2}+s$)
symmetry.  To reconcile the various results, M{\"u}ller proposed
that the surface might be mostly
$d_{x^2-y^2}$-wave, and the bulk  mostly  $s$-wave
\cite{mueller}. Especially in
Bi$_2$Sr$_2$CaCu$_2$O$_{8+\delta}$ (BSCCO), the $c$-axis
transport above $T_c$ is incoherent, 
\cite{timusk}, and scanning tunneling microscope studies revealed that it is   electronically disordered on the scale of the  superconducting
coherence length, $\approx$1.5 nm \cite{lang}, both unfavorable features 
for bulk $d$-wave superconductivity.  Both $c$-axis Pb/BSCCO Josephson junctions
\cite{mk} and a new phase-sensitive
experiment  on BSCCO are
consistent with those observations \cite{Li}.

\section{The Bicrystal Twist Experiment}
Li {\it et al.} cleaved a single crystal of BSCCO, twisted the two  cleaves an
angle $\phi_0$ about the $c$-axis and fused them together, forming junctions of
remarkably superior quality \cite{Li}. High resolution transmission electron
spectroscopy and other studies revealed that the junction cross-sections were atomically
clean over more than 5 $\mu$m \cite{zhu}, far superior to those used in  tricrystal
experiments \cite{tk}. They  measured
the $c$-axis critical current $I^S_c$ and $I^J_c$ across a single crystal and
the twist junction near to $T_c$, respectively, and the respective areas $A^S$ and
$A^J$.  They found that the ratio of the critical current densities
$J^J_c=I^J_c/A^J$ to $J^S_c=I_c^S/A^S$ at $0.9T_c$ was independent of $\phi_0$, as
shown in Fig. 1 \cite{Li}.  
Here we argue that  these data demonstrate that the bulk OP in BSCCO is
$s$-wave for $T\le T_c$, and the $c$-axis quasiparticle tunneling is strongly incoherent.

\begin{figure}[b]
\vspace*{-3mm}
\epsfxsize=8.5cm
\centerline{\epsffile{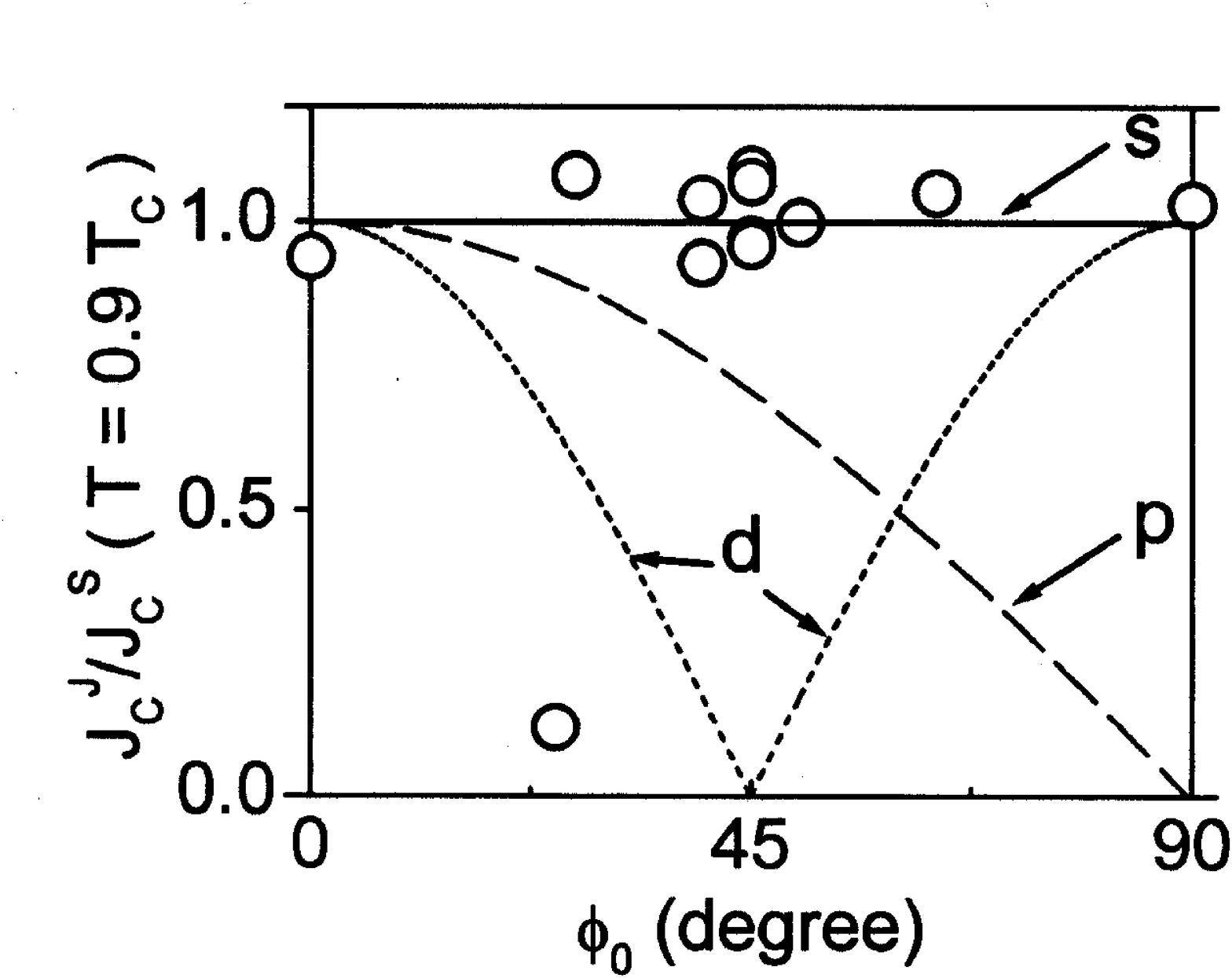}}
\vspace*{-3mm}
\caption{Ratio at $0.9T_c$ of the critical current densities $J^J_c/J^S_c$ across $c$-axis
twist junctions to that across single crystals of BSCCO, versus the
twist angle $\phi_0$. [6] The curves 
are theoretical results for   strongly 
incoherent
$c$-axis tunneling. [10] }\label{fig1}
\end{figure}

\section{Group Theory and the Fermi Surface}
Both YBCO and BSCCO are orthorhombic, but in different ways.  
For YBCO with point group $C_{2v}^1$, the mirror
planes $\sigma_x, \sigma_y$ (the $ac, bc$ planes) contain the  crystal axes $a$ and $b$ along the
Cu-O bond direction in the CuO$_2$ layers.  In BSCCO with approximate point
group $C_{2v}^{13}$, the mirror plane $\sigma_b$ (the $bc$-plane) contains the  $b$ crystal axis (along a diagonal
between the Cu-O bond directions) and the periodic
lattice distortion \cite{zhu,krs}.  The irreducible
representations for the OPs in YBCO and BSCCO are  given in Table I.
Although $s$- and $d_{x^2-y^2}$-wave OP components are compatible in
the bulk of YBCO, they  are {\it incompatible} in  BSCCO, requiring
a second phase transition for bulk coexistence. 

We assume the  quasiparticle dispersion has either the tight-binding   $\xi({\bf k})=
-t[\cos(k_xa)+\cos(k_ya)]+t'\cos(k_xa)\cos(k_ya)-\mu$ with $t = 306$ meV,
$t'/t=0.90$, and $\mu/t=-0.675$, or  hot spot
$[\cos(k_xa)-\cos(k_ya)]^2-\nu^2$ forms,  and a respective tetragonal Fermi
surface (FS) with $\xi({\bf k}_F)=0$, shown in Fig. 2 \cite{ak,bks}.

\begin{figure}[t]
\epsfxsize=8.5cm
\centerline{\epsffile{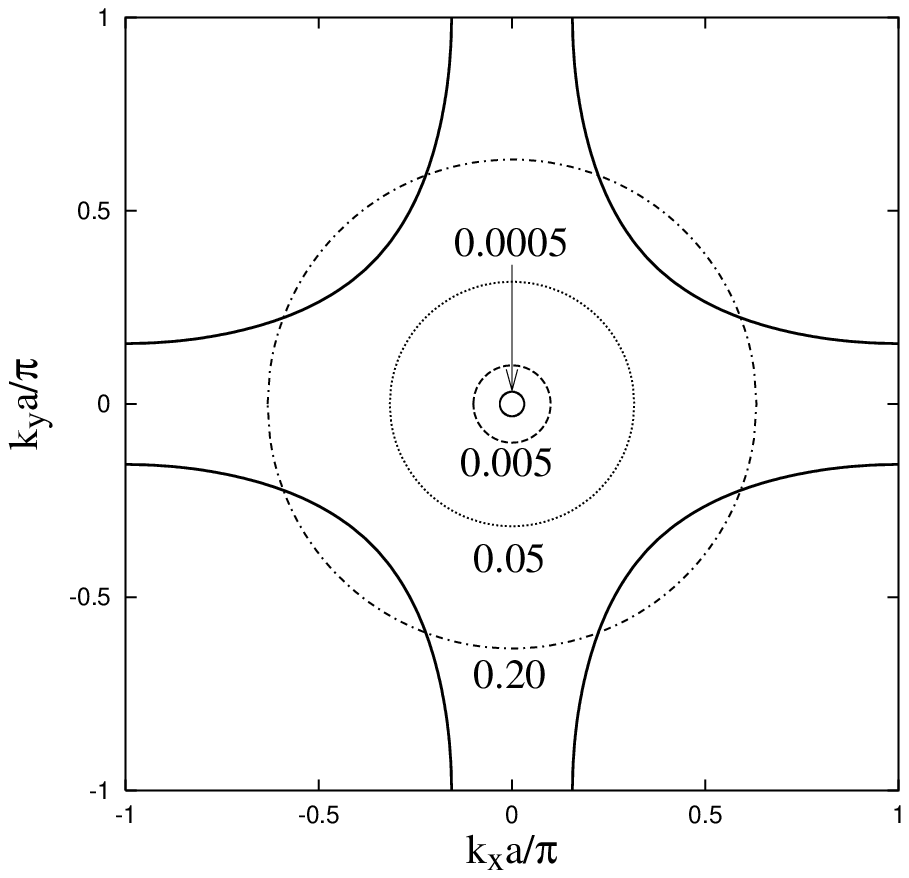}\hspace*{-20pt}}
\vspace*{3mm}
\epsfxsize=8.5cm
\centerline{\epsffile{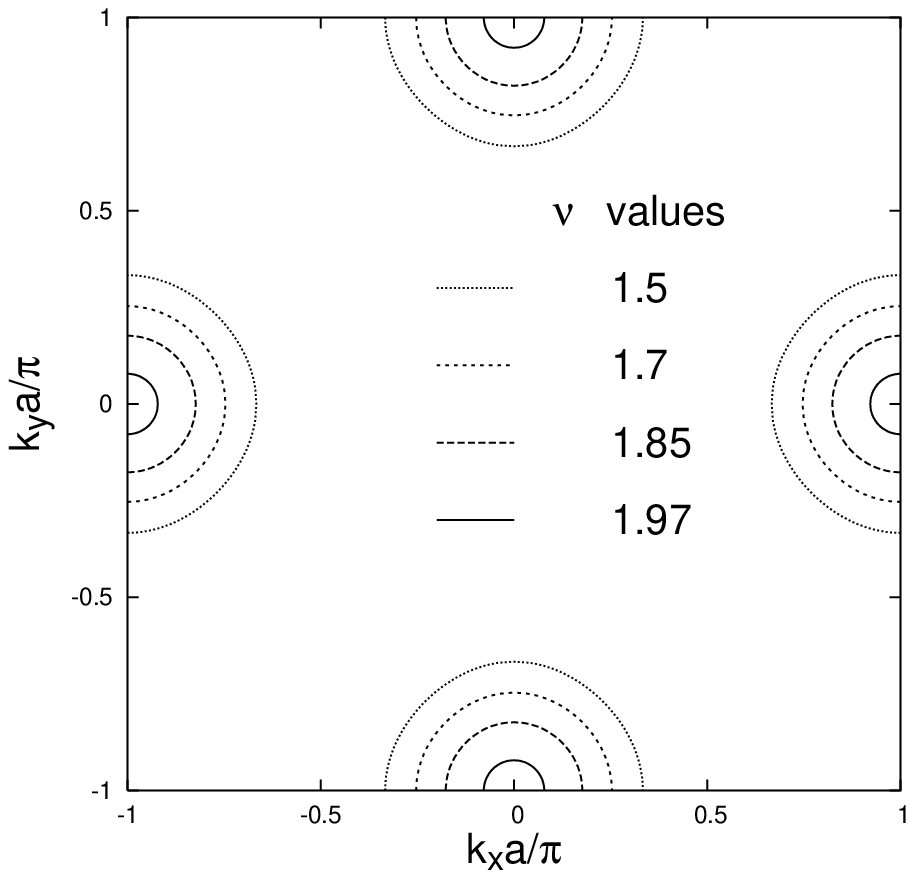}\hspace*{-20pt}}
\caption{Upper: tight-binding FS (solid) plus $f^J$
half-width regions for  indicated $\tilde{\sigma}^2$ values. [10] Lower:  hot spot FSs with $\nu= 1.5, 1.7, 1.85, 1.97$.}\label{fig2}
\end{figure}

\section{The Twist Theorem and its Consequences}
For weak tunneling across the twist junction,  
\begin{eqnarray}
J_c^J(\phi_0)&=&|4eT\sum_{\omega}\langle f^J_{{\bf k},{\bf
k'}}F_{\omega,{\bf k}}[R_{{\bf k}'}(\phi_0)F_{\omega,{\bf k}'}]\rangle|,
\end{eqnarray}

\begin{table}[b]
\caption{Irreducible representations (IR) of the OPs for orthorhombic point groups
$C_{2v}^1$ for YBCO (left) and $C_{2v}^{13}$ for BSCCO (right). }
\begin{tabular}{lll|lll}\hline
IR & YBCO OP & $\sigma_x,\sigma_y$ &IR & BSCCO OP & $\sigma_b$\\ \hline
$A_1$ &  $|s+d_{x^2-y^2}\rangle$ & +1 & $A'_1$ & $|s+d_{xy}\rangle$ & +1\\
$A_2$ & $|d_{xy}+g_{xy(x^2-y^2)}\rangle$ & -1 & $A'_2$ &
$|d_{x^2-y^2}+g_{xy(x^2-y^2)}\rangle$ & -1\\
\hline
\end{tabular}
\end{table}

where $F_{\omega,{\bf k}}=\Delta({\bf k})/[\omega^2+\xi^2({\bf k})+|\Delta({\bf
k})|^2]$, $\Delta({\bf k})$ is the OP, $\omega$ represents the Matsubara
frequencies, $f^J_{{\bf k},{\bf k}'}$ is the tunneling matrix element squared,
$\langle\ldots\rangle$ is an average over each first Brillouin zone (BZ), and
$R_{{\bf k}'}(\phi_0)$ rotates the wave vectors ${\bf k}'$ by
$\phi_0$ about the $c$-axis.   For a single $d$-wave OP component, $R_{\bf
k}(\pi/2)\Delta({\bf k})=-\Delta({\bf k})$.

\begin{twist}
For any weak tunneling matrix element squared satisfying $f^J_{{\bf k},{\bf
k}'}=f^J_{{\bf k}',{\bf k}}$, an arbitrary OP of general $d_{x^2-y^2}$- or
$d_{xy}$-wave symmetry in a tetragonal crystal gives rise to a vanishing
$c$-axis critical current across an internal 45$^{\circ}$ twist junction for
$T\le T_c$.
\end{twist}
\vskip0pt
\noindent{\bf PROOF}
\vskip0pt
\begin{eqnarray}
Z_{\omega}&=&\langle f^J_{{\bf k},{\bf k}'}[R_{{\bf k}'}(\pi/4)F_{\omega,{\bf
k}'}]F_{\omega,{\bf k}}\rangle\nonumber\\
& =&\langle f^J_{{\bf k},{\bf k}'}F_{\omega,{\bf
k}'}[R_{{\bf k}}(-\pi/4)F_{\omega,{\bf k}}]\rangle\nonumber\\
&=&\langle f^J_{{\bf k},{\bf k}'}F_{\omega,{\bf
k}'}[-R_{{\bf k}}(\pi/4)F_{\omega,{\bf k}}]\rangle\nonumber\\
&=&\langle f^J_{{\bf k}',{\bf k}}F_{\omega,{\bf
k}}[-R_{{\bf k}'}(\pi/4)F_{\omega,{\bf k}'}]\rangle=-Z_{\omega}=0.\nonumber
\end{eqnarray}

\begin{figure}[t]
\epsfxsize=8.5cm
\centerline{\epsffile{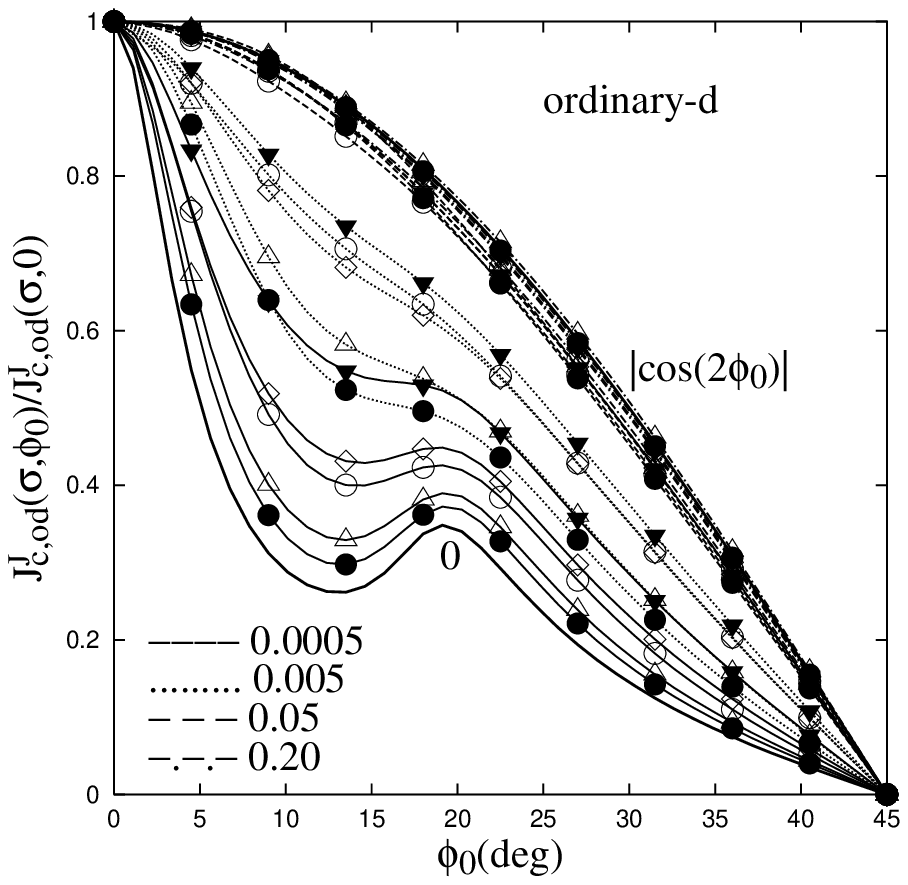}\hspace*{-20pt}}\vspace*{5pt}
\epsfxsize=8.5cm
\centerline{\epsffile{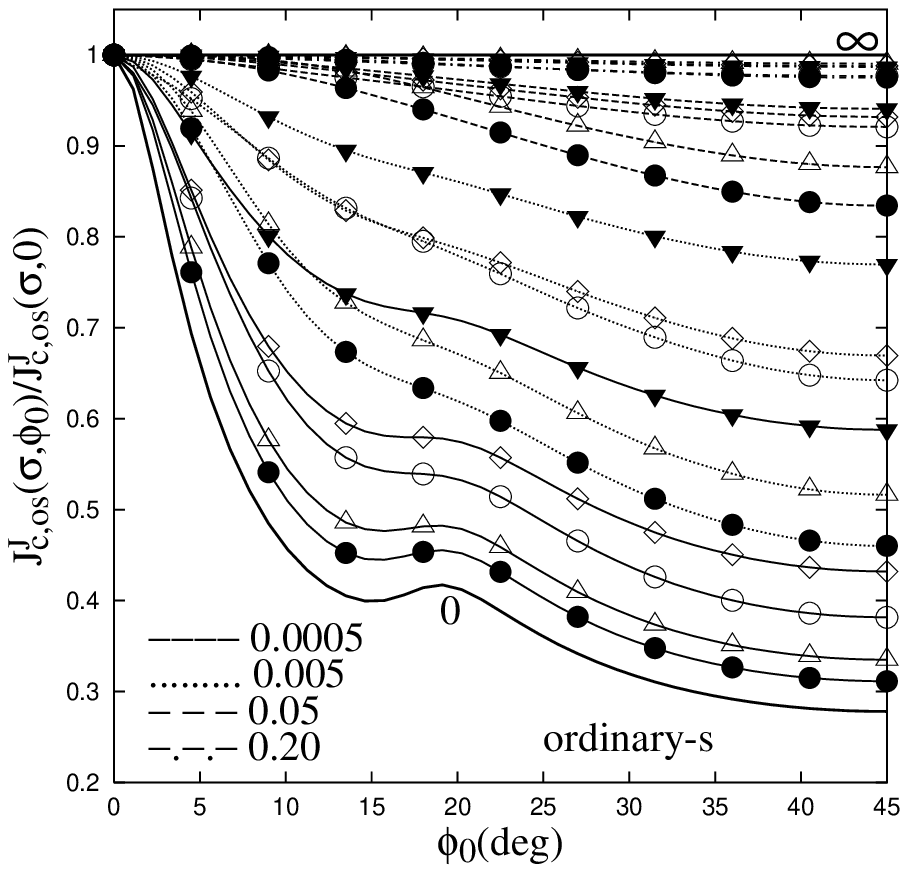}\hspace*{-20pt}}
\caption{Plots of $J_c^J(\phi_0)/J_c^J(0)$ near $T_c$ with the tight-binding
FS in Fig. 2 for the ordinary $d_{x^2-y^2}$ (top)
and $s$ (bottom) OPs.  The thick solid lines labelled 0 are for coherent tunneling,
and those labelled  $|\cos(2\phi_0)|$ and $\infty$ are for
purely incoherent tunneling.   Other curve types labelled with the values of 
$\tilde{\sigma}^2$ measure the fraction of the first BZ
involved in the tunneling, as shown in Fig. 2. Results for the  Gaussian ($\bullet$), exponential ($\circ$), Lorentzian ($\diamond$)  rotationally-invariant
Lorentzian ($\triangle$), and stretched Lorentzian (solid inverted triangles)
 $f^J$ forms are shown. [10] }\label{fig3}
\end{figure}
\begin{figure}[floatfix]
\vspace*{-3mm}
\epsfxsize=8.5cm
\centerline{\epsffile{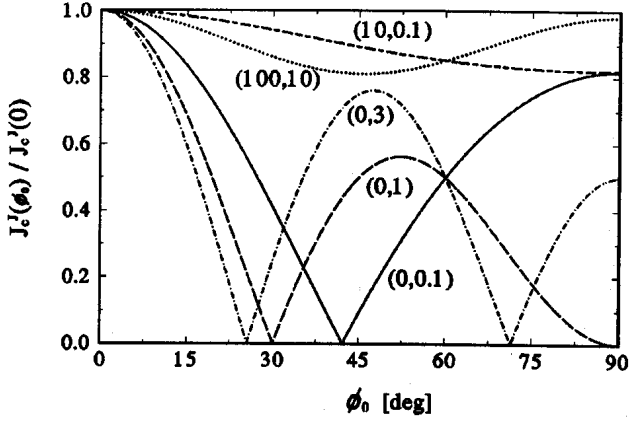}}
\caption{Plots of $|A+\cos(2\phi_0)+B\cos(4\phi_0)|/|1+A+B|$, with the
$A,B$ values, for $J_c^J(\phi_0)/J_c^J(0)$ from the $A_1', A_2'$ OP IRs.
[8]}\label{fig4}
\end{figure}

We studied a variety of OP and weak $f^J_{{\bf k},{\bf k}'}$ forms \cite{bks}.  In Fig. 3, 
we show our
results for the
ordinary-$d$- and ordinary-$s$- OP forms proportional to
$\cos(k_xa)-\cos(k_ya)$ and $1$, respectively. For a Gaussian
form of $f^J$, $f^J_{{\bf
k},{\bf k}'}=f^J_0\exp[-({\bf k}-{\bf k}')^2/\tilde{\sigma}^2]$.  We also
studied exponential, Lorentzian, rotationally-invariant Lorentzian, and
stretched Lorentzian forms \cite{bks}.  
  Regardless of the form of $f^J$, the twist theorem
requires $J_c^J(\pi/4)=0$ for a $d$-wave OP. From these and similar unpictured curves, it is evident
that only an  OP of general $s$-wave symmetry can fit the data.  Moreover,
 the $c$-axis quasiparticle tunneling must be very incoherent.  

\subsection{Orthorhombicity}

When orthorhombicity is included, the theorem is not rigorous.  However, 
including a small $g_{xy(x^2-y^2)}$ OP component in a dominant-$d_{x^2-y^2}$
$A_2'$ 
OP will only shift the angle $\phi_0^*$ at which $J_c^J(\phi_0^*)=0$  by a small amount from 45$^{\circ}$, as
pictured in Fig. 4.  Hence, orthorhombicity cannot explain 
the data of Li {\it et al.} \cite{Li}.
\subsection{Order Parameter Twisting}

At low $T$, it is possible to obtain  $J_c^J(\pi/4)\ne0$ with a
predominant-$d_{x^2-y^2}$ $A_2'$ OP symmetry, provided that a subdominant
$A_1'$ OP component can
exist. Near to the twist junction,  the dominant
$A_2'$ OP would be suppressed, and the subdominant $A_1'$ OP 
increases in amplitude, so that the overall OP effectively rotates near the
twist junction \cite{krs}.  However, the amount of twisting
is strongly limited by the second bare transition temperature $T_{cB}^0$ and
by the bulk and twist junction Josephson coupling strengths $\eta, \eta'$.  In Fig. 5, we show the
results obtained with subdominant OPs of the $d_{xy}$- and $s$-wave forms, respectively.  Neither
 case can fit the data of Li {\it et al.} \cite{Li}.  Figure 5 shows
that an experiment just below $T_c$ can rule out OP twisting
effects, unless $T_{cB}^0=T_c$, for which the overall bulk OP would be 
nearly isotropic at low $T$.
\begin{figure}[t]
\epsfxsize=8.5cm
\centerline{\epsffile{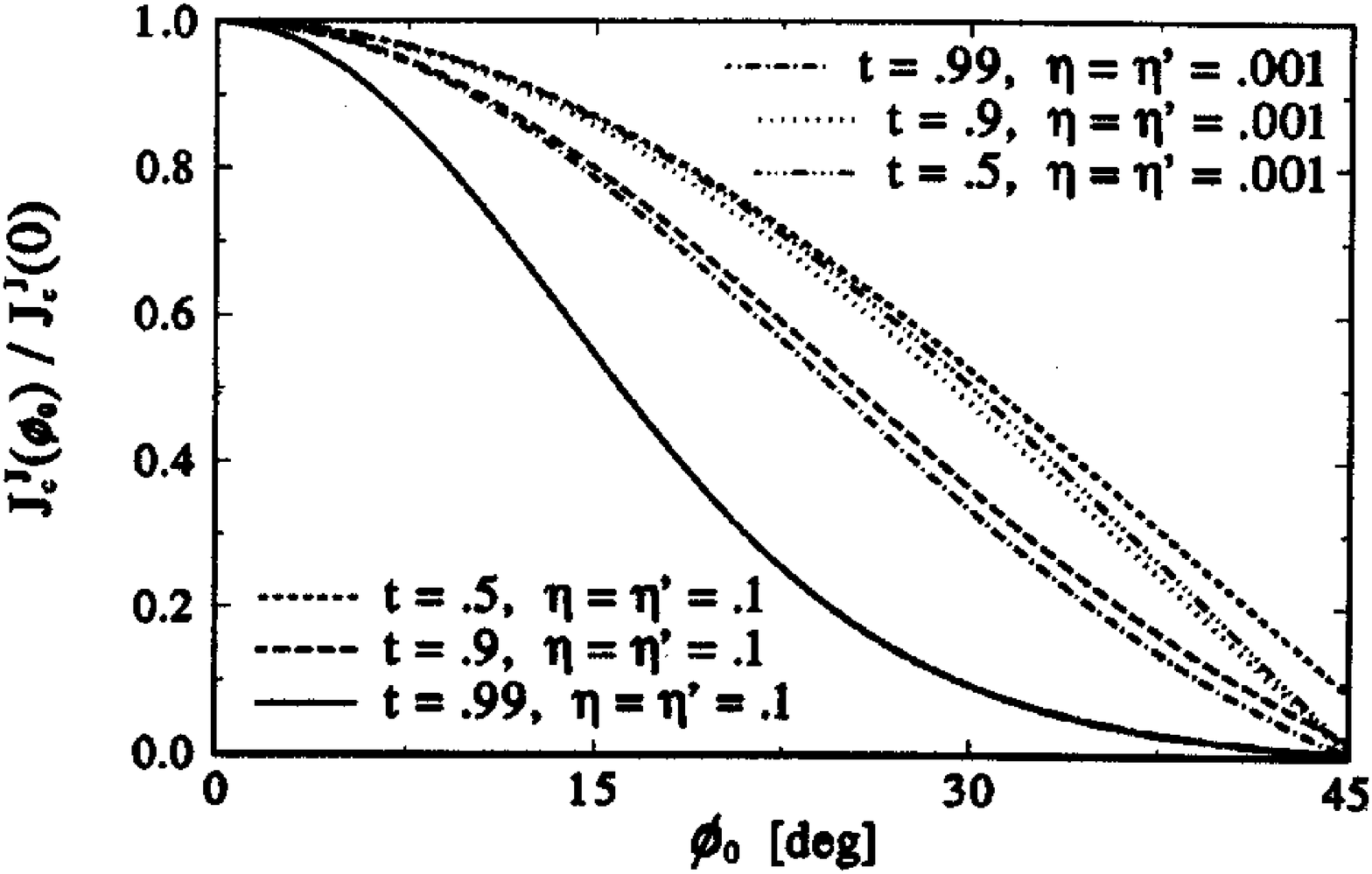}}
\epsfxsize=8.5cm
\centerline{\epsffile{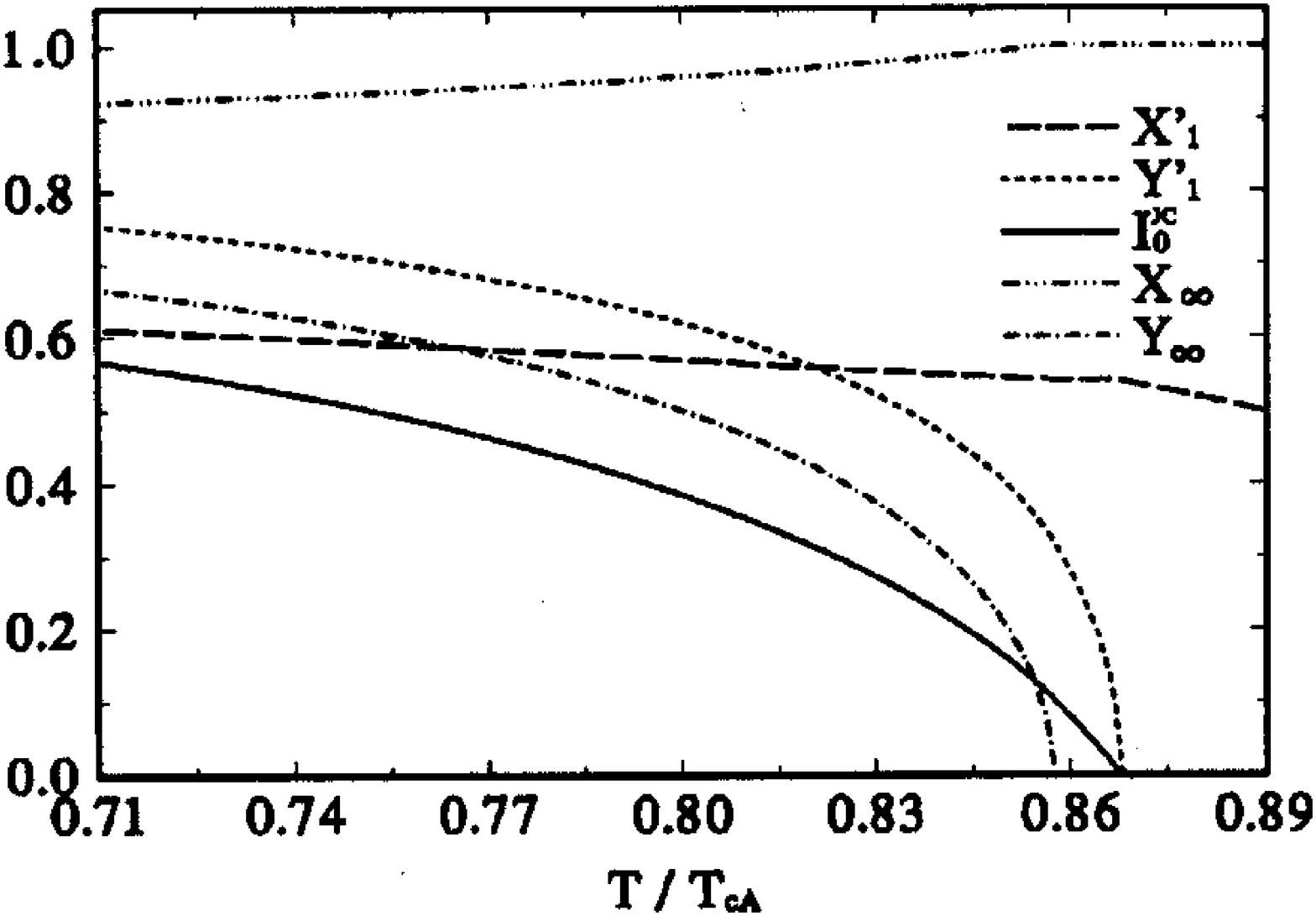}\hspace*{-5pt}}
\caption{Upper:  Plots of $J_c^J(\phi_0)/J_c^J(0)$ at different $t=T/T_c$
values for a  $d_{x^2-y^2}+id_{xy}$ symmetry in the 
bulk, with $T_{cB}^0=0.1T_c$. $\eta$ and $\eta'$ are the
Josephson couplings across bulk and the twist junctions, respectively. Lower (solid):
$T/T_{cA}=t$ dependence of $J_c^J(45^{\circ})$ for a twist junction with 
$d_{x^2-y^2}+is$ symmetry in the bulk, and $T_{cB}^0=0.9T_c$. [8]}\label{fig5}
\end{figure}
\subsection{Strong Coupling and Other Models}

We also considered both weak and strong coherent $c$-axis tunneling with a
variety of FS forms \cite{ak}. In Fig. 6, we show the tight-binding FS $J_c^J(\phi_0)$ for second order
coherent tunneling.  In Fig. 7, the $J_c^J(\phi_0)$ for weak coherent tunneling with two hot spot FSs in
Fig. 2 are shown.  None of these curves
fit the data of Li {\it et al.}\cite{Li}.  
\section{BSCCO Whisker $c$-Axis Twist Experiments}

Recently, Takano {\it et al.} performed low-$T$
  $c$-axis twist mesa experiments using overdoped BSCCO whiskers with $45^{\circ}\le\alpha\equiv\phi_0\le90^{\circ}$
\cite{takano}.  Their data, pictured in Fig. 7,  are distinctly different from
those of Li {\it et al.} \cite{Li}, with a strong $J_c^J(\phi_0)$ dependence. 
Especially since $J_c^J(\phi_0)$ for $\phi>80^{\circ}$ was anomalously large,
Takano {\it et al.} suspected an
extrinsic $\phi_0$ dependence to $f^J_{\bf k,{\bf k}'}$
\cite{takano}. Nevertheless, in Fig. 7 we  fit the data
using Eq. (1) by assuming the
quasiparticles have a hot
spot dispersion and intrinsically coherent $c$-axis tunneling.
  Subsequently, they found $J_c^J(\phi_0)\approx C\ne0$ from many junctions
with $\phi_0\approx 45^{\circ}$, and provided
preliminary Fraunhofer and Shapiro evidence that the non-vanishing
$J_c^J(45^{\circ})$ arises from  first-order Josephson tunneling \cite{tachiki}.  Hence, the whisker
experiments rule out a pure $A_2'$ (e. g., $d_{x^2-y^2}$) OP, but  are presently consistent
with an OP either of pure $A_1'$ (e. g., $s$) symmetry, or of mixed 
$A_1'$ and $A_2'$ (e. g., $d_{x^2-y^2}+is$)  symmetry.  From our theoretical
studies, whisker experiments just below
$T_c$ might determine if their bulk OP is also 
pure $A_1'$ \cite{Li}, and  measurements above $T_c$
could investigate if the $c$-axis transport is indeed coherent, strikingly inconsistent
with single crystal BSCCO
\cite{timusk}.

\section{Conclusions}

The  data of Li {\it et al.} demonstrate that the OP in the bulk of
BSCCO has $A_1'$ ($s$) symmetry for $T\le T_c$, and that the $c$-axis tunneling is strongly
incoherent.  The  data of Takano {\it et al.} presently rule out a
pure $A_2'$ ($d_{x^2-y^2}$) OP, but surprisingly  suggest that the weak $c$-axis
tunneling in BSCCO whiskers might be  coherent with a  FS consisting of small hot spots.
 
\begin{figure}[t]
\epsfxsize=7.5cm
\centerline{\epsffile{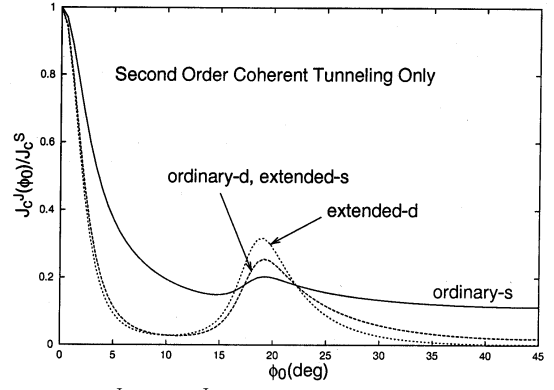}}
\caption{$J_c^J(\phi_0)/J_c^J(0)$ near to $T_c$ for the $s$, $d_{x^2-y^2}$, 
extended-$s$ [$|\cos(k_xa)-\cos(k_ya)|$] and extended-$d$ OPs, [10]
obtained for coherent second order twist junction tunneling  only.}\label{fig6} 
\end{figure}
\begin{figure}[t]
\epsfxsize=8.5cm
\centerline{\epsffile{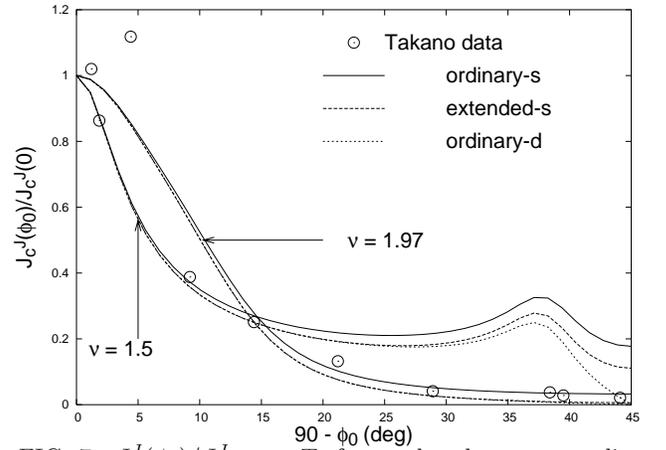}\hspace*{-10pt}}
\caption{$J_c^J(\phi_0)/J_c^J$ near $T_c$ for weak coherent tunneling with
the hot spot FSs in Fig. 2 with $\nu=1.5, 1.97$, for the ordinary-$s$, extended-$s$,
and ordinary-$d_{x^2-y^2}$ OPs.  The  whisker data ($\circ$)  of Takano {\it et
al.} are also shown. [11] }\label{fig7} 
\end{figure}
\section*{acknowledgments}
We thank G. B.  Arnold, Qiang Li, K. A. M{\"u}ller, K. Scharnberg, and
M. Tachiki for useful discussions.

%
%

\end{document}